# Immersive and Collaborative Data Visualization Using Virtual Reality Platforms

Ciro Donalek, S. G. Djorgovski, Alex Cioc, Anwell Wang,
Jerry Zhang, Elizabeth Lawler, Stacy Yeh,
Ashish Mahabal, Matthew Graham, Andrew Drake
California Institute of Technology
Pasadena, CA 91125, USA
[donalek,george]@astro.caltech.edu

Scott Davidoff, Jeffrey S. Norris
Jet Propulsion Laboratory
Pasadena, CA 91109, USA
[Scott.Davidoff, jeffrey.s.norris]@jpl.nasa.gov

Giuseppe Longo
University Federico II
Napoli, Italy
longo@na.infn.it

*Abstract* — **Effective data visualization is a key part of the discovery process in the era of "big data". It is the bridge between the quantitative content of the data and human intuition, and thus an essential component of the scientific path from data into knowledge and understanding. Visualization is also essential in the data mining process, directing the choice of the applicable algorithms, and in helping to identify and remove bad data from the analysis. However, a high complexity or a high dimensionality of modern data sets represents a critical obstacle. How do we visualize interesting structures and patterns that may exist in hyper-dimensional data spaces? A better understanding of how we can perceive and interact with multi-dimensional information poses some deep questions in the field of cognition technology and human-computer interaction. To this effect, we are exploring the use of immersive virtual reality platforms for scientific data visualization, both as software and inexpensive commodity hardware. These potentially powerful and innovative tools for multi-dimensional data visualization can also provide an easy and natural path to a collaborative data visualization and exploration, where scientists can interact with their data and their colleagues in the same visual space. Immersion provides benefits beyond the traditional "desktop" visualization tools: it leads to a demonstrably better perception of a datascape geometry, more intuitive data understanding, and a better retention of the perceived relationships in the data.**

I. INTRODUCTION

Challenges of the "big data" in terms of the data rates and volumes are well known. Even more interesting than the growth of data volumes is the dramatic increase in data complexity: multi-dimensional data sets often combine numerical measurements, images, spectra, time series, categorical labels, text, etc. Feature vectors with tens, hundreds, or even thousands of dimensions at once encapsulate the key challenge of data-driven discovery: the scope of data and processing introduce new opportunities to find connections in data, while at the same time they demand a new set of tools to support the particular qualities of investigating massive data.

*The key point is that "big data science" is not about data: it is about discovery and understanding of meaningful patterns hidden in the data.* Massive and complex data sets – no matter how content-rich or how expensively obtained – are of no use if we cannot discover interesting patterns in them. This applies to data both from measurements, and data generated as models or simulations that need to be compared and interpreted.

Visualization is the main bridge between the quantitative content of the data and human intuition, and it can be argued that we cannot really understand or intuitively comprehend anything (including mathematical constructs) that we cannot visualize in some way. Humans have a remarkable pattern recognition system in our heads, and the ability for knowledge discovery in data-driven science depends critically on our ability to perform effective and flexible visual exploration. This may be one of the key methodological challenges for the data-rich science in the 21st century.

Different data analysis methods, tools, or approaches also depend critically on the geometry and topology of data distribution in the relevant parameter space. A blind application of algorithms based of erroneous assumptions inevitably leads to wrong or misleading results. Visualization thus must be an integral part of any data mining process. It is also essential in the data preparation process, as missing or potentially anomalous measurements can be readily identified by inspection, and removed from the analysis.

A high complexity or a high dimensionality of modern data sets thus represents a critical obstacle: we are biologically optimized to see the world and the patterns in it in 3 dimensions. How do we visualize structures that may exist in hyper-dimensional data spaces? For example, there may exist clustering in tens of dimensions, or multivariate correlations embedded in higher dimensionality manifolds. In general, some low-dimensionality projection of high-dimensional structures would smear out the structures that may be present in the data, and thus render them effectively unrecognizable. Various dimensionality reduction methods do exist, but may not always be applicable. The more dimensions we can visualize effectively, the higher are our chances of recognizing potentially interesting patterns, correlations, or outliers.

This paper is a progress report on our initial exploration of the utility and optimal use practices of immersive Virtual Reality (VR) as a platform for an interactive, collaborative, scientific data visualization and visual exploration.

## II. VIRTUAL REALITY AS VISUALIZATION PLATFORM

VR has been shown to lead to better discovery in domains that whose primary dimensions are spatial. Early work demonstrated that immersion helps scientists more effectively investigate paleontology [1], brain tumors [2], shape perception [3], underground cave analysis structures [4], MRI [5], organic chemistry [6] and physics [7]. Data visualization has been shown to support highly abstract multi-dimensional analyses. Many researchers look to visualization to support exploration of large datasets.

One of our goals is to investigate the confluence of immersive virtual reality and abstract visualization, and the interplay between novel technology and human perception to generate insight into both. VR and abstract data visualization have each independently been demonstrated to improve science outcomes, but to the best of our knowledge, there is no evaluation of how immersive virtual reality can use abstract representation of high-dimensional data in support of scientific investigation. Because immersive visualization can multiply the effectiveness of desktop visualization, we propose that immersive visualization become one of the foundations to explore the higher dimensionality and abstraction that are attendant with "big data".

Separately, VR is also proving it can support collaborative tasks. When used as a medium for telepresence [8] – the feeling of being there – VR has been shown to increase situation awareness [9] and the vividness [10], interactivity [8] and media richness [11], as well as the proprioceptive [12] or kinesthetic [13] properties of remote experiences. These qualities have been shown to enhance collaborative teleoperation scenarios like remote driving [14] and surgery [15], just to cite a few. However, so far there has been no evaluation of how immersive VR can use abstract representation of high-dimensional data and feature spaces in support collaborative investigation of scientific data.

Whereas the studies cited represent some very good use cases, they tend to be very case-specific (or one-of), and we know of no VR system that has looked to exploit the general purpose visual and interactive problem-solving of abstract, multi-dimensional data. In other words, use of immersive VR as a general purpose tool for data exploration.

Traditional approaches involve use of very complex, expensive, and non-portable equipment such as "caves", hyperwalls, etc. Our goal is to develop powerful, flexible, portable data visualization tools that can be used on a standard, affordable desktop or laptop, possibly supplemented with inexpensive commercially developed hardware (e.g., *Oculus Rift*[TM] headset, *Leap Motion*[TM] "3D mouse", *Microsoft Kinect*[TM] sensor), that can be competitive with vastly more expensive cave-type visualization facilities on analytic power and flexibility, and also an incomparable portability.

Our initial focus is on an effective visualization of highly dimensional data that can be represented as feature vectors (or data points) in some high dimensionality data parameter space. We plan to address visualization of non-discrete or continuum variables or multidimensional density fields in the future.

We also leverage the substantial commercial software development for the virtual environments such as video games or virtual worlds (VWs), with the associated engines/libraries, and focus on the development of scientific data visualization tools operating within such environments. At the same time, these platforms provide an easy and natural path to a collaborative data visualization and exploration, where scientists can interact with their data and their colleagues in the same visual space.

## III. SOME PRELIMINARY DEVELOPMENTS

As a first step towards addressing these challenges, we conducted some preliminary inquiries and technical demonstrations as described below.

We conducted an experiment in the possible scientific and scholarly uses of virtual worlds (which are general purpose immersive VR platforms) such as *Second Life*[TM] (SL) and *OpenSimulator* (OpenSim), an open source version of the SL software. The initial experiments were done through the Meta-Institute for Computational Astrophysics (MICA), the first professional scientific organization based in VWs [16, 17, 26], subsequently moving to an *OpenSim*-based VW, *vCaltech*.

Use of such "off the shelf" VWs offers several advantages. First, all of the graphical rendering, geometry and interactivity issues are already solved; all that needs to be done is scripting for a specific purpose of data visualization. They come with suitable software libraries and building tools. Their cost is zero and user interaction is built in.

We have developed scripts for a general, immersive visualization of highly dimensional data sets in VWs, using the SL Linden Scripting Language (LSL), which has a structure that is largely based on Java and C, and its *OpenSim* dialect; *OpenSim* also uses C# for some functions.

As our test data sets, we used catalogs of objects detected in large digital sky surveys, e.g., [18, 19, 20], which can be represented as feature vectors in a space of tens to hundreds of dimensions. These may include observed brightness of the sources, quantitative measures of their spectral light distributions, expressed as flux ratios in different filters (or, in the astronomical parlance, colors), various measures of their spatial extent and the morphology of the light distribution on the sky (e.g., unresolved, point sources, vs. resolved galaxies of various morphologies), physical properties derived from these and other measurements, e.g., luminosities, etc. A challenge that often arises in the analysis of such data is an automated classification of different types of objects [21, 22, 23].

where different users can interact by voice and/or text in the same visual space where the data are being rendered.

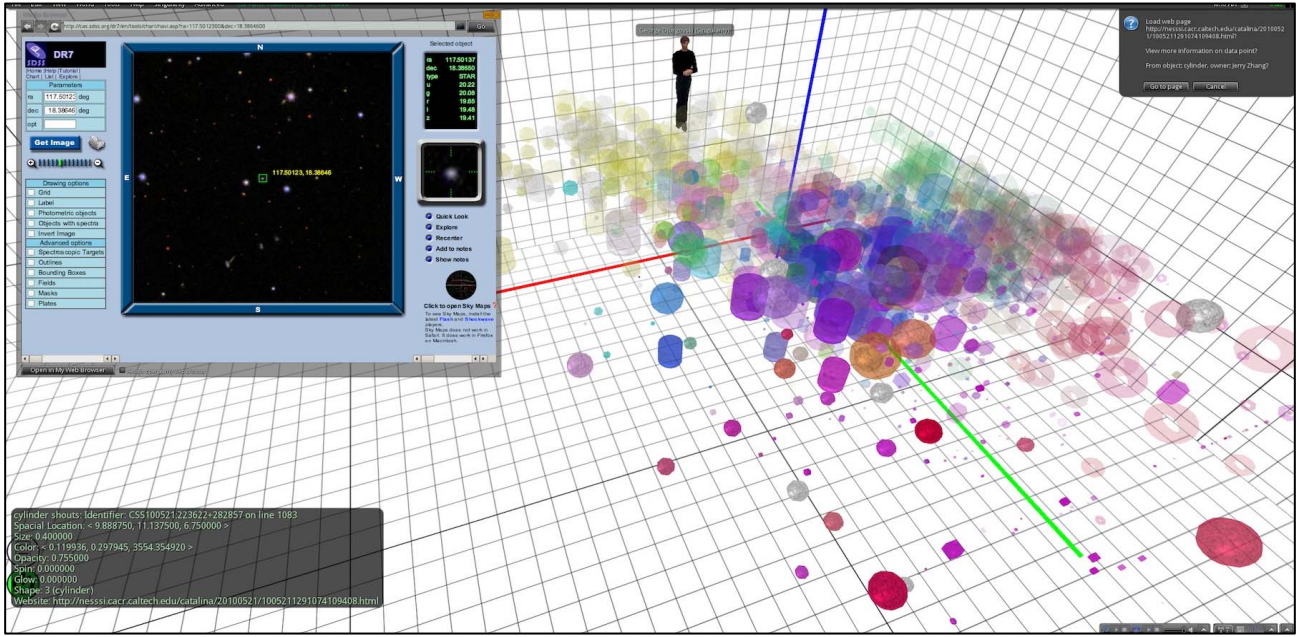

Fig. 1. A student, represented by his avatar near the top of the image, performing data visualization experiments in the *OpenSim*-based virtual world *vCaltech*. Different data parameter values are mapped into the displayed XYZ, data point shapes, sizes, colors, and transparencies, effectively representing an 8-dimensional data visualization. Here we added the ability to embed links in the data point that would bring up a webpage with additional data for a given object from an external database, which aids the interpretation of visually observed patterns or outliers. (We note that these flat figures – screen grabs – do not convey by far the quality of the user interaction in these immersive VR environments.)

We experimented with different approaches of such multidimensional data representation, e.g., the XYZ spatial coordinates (in some data parameter space, not the physical space), colors, sizes, transparencies and shapes of data points, their textures, orientations, rotation, pulsation, etc., to encode a maximum number of data dimensions (Fig. 1).

In the *OpenSim* based vW, *vCaltech*, we used the so-called region modules, custom codes that can greatly accelerate visualization and allow to easily visualize $\sim 10^4 - 10^5$ data objects ("prims").

One implemented functionality is the ability to link data objects with external catalog or database information, e.g., using a simple point-and-click. This allows the user to connect directly the immersive 3D data display with the additional archival information that cannot be encoded in the visual display, as well as the hyperlinks that can be used to extend the data exploration, and potentially lead to a physical interpretation of the visually observed patterns in the data.

Since VWs are designed around the notion of multiple user interactions in immersive 3D spaces, this approach directly enables a collaborative data visualization and exploration,

IV. iVIZ: A NEW, PRACTICAL DATA VISUALIZATION TOOL

While "off the shelf" VWs have several advantages, they are not optimized for an efficient rendering of massive data sets, have relatively limited scripting functionalities, and require use of separate browsers. Thus, we have started a development of data visualization tools using the *Unity 3D*™ platform, which is now emerging as a dominant immersive VR platform (mostly for the game development, but also for the professional activities). We can already easily and rapidly visualize $10^5$ to $10^6$ data points (Fig. 2), which is comparable to other state of the art visualization tools.

The prototype data visualizer based on the Unity 3D, with a working name iViz (Fig. 2), is multiplatform and can run as a standalone application or in a web-browser. It also allowed us to develop additional functionalities that are not easily doable on the OpenSim platform. Using *Unity 3D* has an important advantage that in addition to a custom visualization browser (as is the case with the SL and *OpenSim*), we can also use a standard web browser as a display. Using a familiar web browser as an interface to an immersive 3D space may facilitate the uptake of this technology by those reluctant to explore the research potential of immersive VR in a superficially "game-like" environment.

We added a new feature for a collaborative, multi-user visual data exploration. In addition to the standard mode of each user having their own viewpoint and navigating through the data independently, we added a broadcasting function where users can have a shared view associated with one user who then navigates through the data. This enables a more flexible, collaborative data exploration [24, 25].

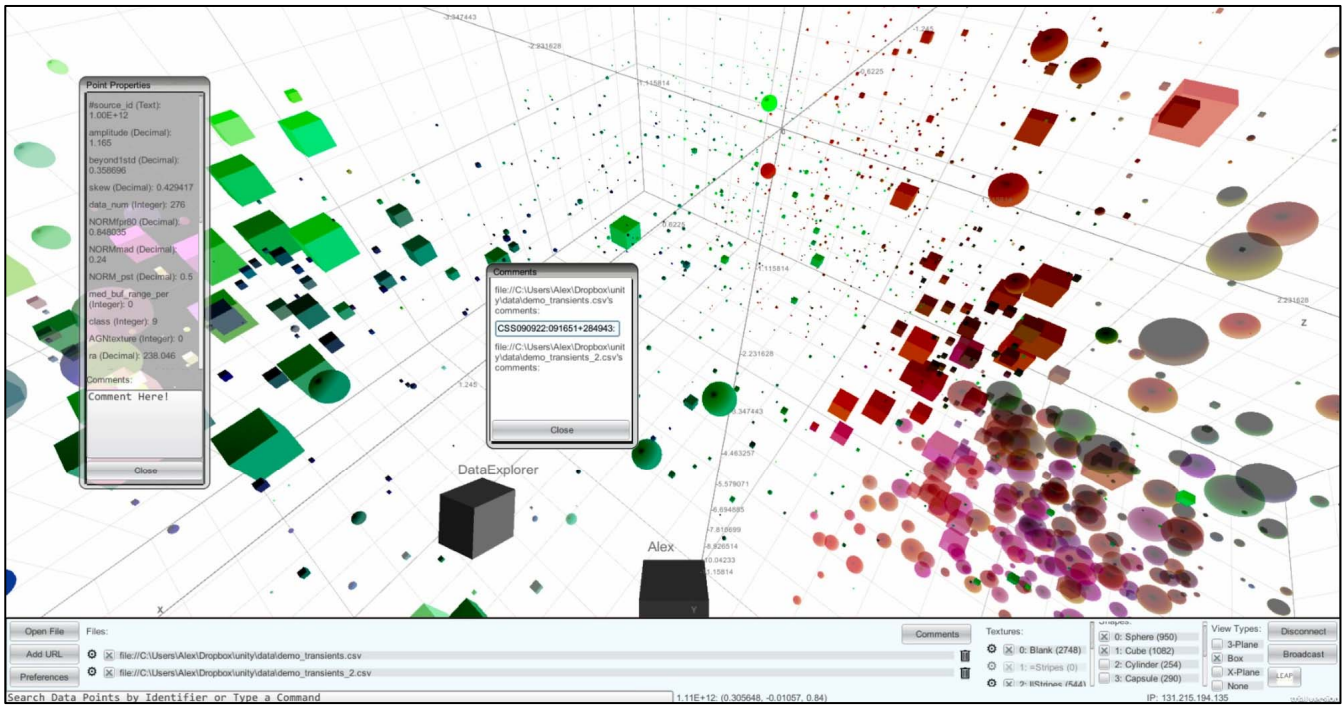

Fig. 2. An example of the current interactive user interface for the *Unity*-based data visualizer, *iViz*. It incorporates the same functionalities from our OpenSim-based immersive visualization, and it adds some new ones, including the ability to change mapping of data axes into the display axes, annotation of data points, and a number of new interaction capabilities.

The user interface for *iViz* enables the user to easily select and shuffle which data parameters are mapped to which graphical "axis" (XYZ positions, colors, shapes, sizes, transparency, textures, etc.), in order to determine the optimal mapping choice for a given scientific application, e.g., clustering of different object classes, a search for outliers, etc. Patterns that are not discernible in one mapping of data dimensions to display "axes" sometimes become obvious with a different mapping. This flexibility enables a more powerful visual data exploration and discovery.

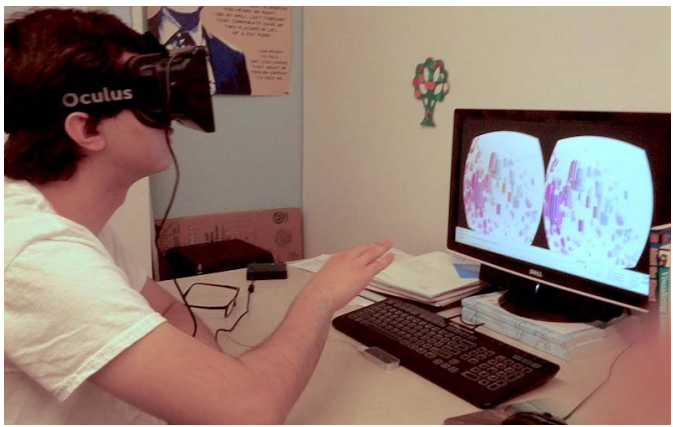

Fig. 3. A student using the *Oculus Rift*[TM] VR goggles and navigating through the visualized data space using the hand motions and the *Leap Motion*[TM] sensor. The computer screen shows the stereoscopic image pair that is being displayed through the VR goggles.

*iViz* fully support the Leap Motion sensor (Fig. 3), effectively a 3D mouse, and Kinect. Both of our *OpenSim* and *Unity* based visualizers already support the *Oculus Rift*[TM] VR goggles as a display device, which greatly enhance the sense of immersion in the data. Since wearing of the goggles effectively precludes the use of a keyboard interface, we need a tool where a user can use physical gestures to manipulate virtual objects and data displays.

V. IMMERSIVE VISUALIZATION OF MARTIAN LANDSCAPE

In order to compare directly the potential advantages of an immersive VR data display over the traditional images, we have used immersive visualization of Martian landscape to give new capabilities to planetary scientists, and tele-robotic operators. Terrain on Earth is available for geologists to walk through, using their highly evolved perceptual capabilities to understand the relative distances between objects, the connection between the near- and foreground, the overall orientation of features, and ultimately to develop theories to explain their geological formation. Scientists studying Mars, however, cannot do that. Instead, they are forced to use various image mosaics.

To understand how scientists make meaning of distant environments mediated by technology, we crafted a controlled study to compare the effects of two technologies. Our main objective was to investigate whether the immersive environment provided scientists a more intuitive understanding of the terrain. Our hypothesis was that scientists viewing a remote environment using immersive VR will have a higher situation awareness than scientists using image panoramas on desktop computers.

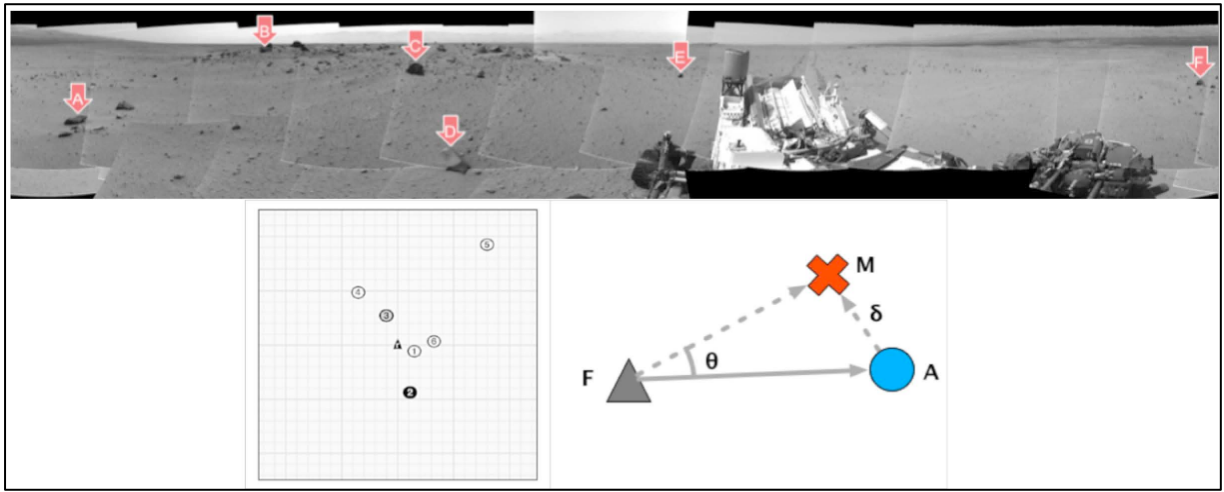

Fig. 4. Top: the panoramic mosaic of the Curiosity Mars rover images used in this experiment, with some of the features of interest indicated with the arrows. Bottom left: an example of a map with the estimated relative positions of different features. Bottom right: a schematic illustration of the geometry used in the computation of distance and angle errors.

We evaluated this hypothesis with a map-drawing task. Since geologists examining formations in an area of interest often begin with mapping the territory, we operationalized situation awareness as the accuracy of drawn maps.

Using a 2×1 between-subjects design, we varied how sensor data was presented to scientists. Scientists in the panorama group were shown a 2-D cylindrical mosaic panorama image stitched from a series of individual snapshots (Fig. 4). Scientists in the immersive VR group were shown a 3D stereoscopic image using an *Oculus Rift*™ head-mounted display (HMD) with a *Vicon*™ motion-capture system, which is typically used for motion capture (e.g., in movies and video games). We used the infrared markers to track the location of the *Oculus Rift*™ in space, thereby tracking the user's position. This gave a user the ability to move in space (in the Vicon volume), turn their head, squat or kneel, and look over and around objects in the scene.

We designed and implemented a map editor, written in *Unity 3D*, to allow for in-situ access to a map creation: when using the HMD, the user should be able to draw the map without leaving the environment. For consistency, the same editor, including the controls, was used when creating the map on the two-dimensional panorama. The editor allowed the user to move the currently-active waypoint, toggle between waypoints, and rotate the map. We used angle and distance errors in a map-making task as a way to quantify situation awareness in the scientific setting. Both distance and angle error were then summed across all landmarks.

Subjects were recruited and selected for their science roles on Mars Exploration Rover (MER) and Mars Science Laboratory (MSL) missions, as scientists (geologists) and science planners. Purposeful snowball sampling was used to ensure diversity of subjects within the role. Subjects were randomly assigned to the immersive condition group or the panorama group. The protocol was identical for both groups. Subjects were trained on the map editor; trained on the system (immersive condition or panorama); presented a scene taken by the Mars Science Laboratory, annotated with six points of interest. Preliminary results are interesting. Anecdotally, scientists seem to be performing better in the immersive condition, but self-reports for each condition are equally and ambiguously positive. For a complete discussion of the experiment, see [27].

## VI. CONCLUDING COMMENTS

Effective data visualization, especially in the context of high-dimensionality parameter or feature spaces, remains as one of the key challenges in the era of "big data" – a cognitive bottleneck on the path between data and discovery. Our goal is to maximize the intrinsic human pattern recognition (or visual discovery) skills through the use of emerging technologies associated with the immersive VR.

Using of these technologies gives us a significant, cost-effective leveraging: their rapid development is paid for by the video gaming and other entertainment industries, and they are steadily becoming more powerful, more ubiquitous, and more affordable. They offer us an opportunity where any scientist can, with a minimal or no cost, have visual data exploration capabilities that are now provided by multi-million dollar cave-type installations, and with a portability of their laptops. Moreover, they open potentially novel ways of scientific interaction and collaboration.

In this preliminary report, we describe our ongoing tools development, and some of the initial experiments. These preliminary studies served multiple goals: (1) they gave us the first insights into the potential of immersive VR as a scientific data visualization platform; (2) they provided us with the first experiences in a collaborative, immersive data visualization, where scientists embedded in a data space can interact both with the data and with their colleagues; and (3) they represent technical proofs of concept, with practical data visualization tools that can be used for a much more detailed and systematic study currently under way, with the results to be reported in a future publication.

In addition to using it ourselves for such studies, we plan to release our *iViz* data visualization platform to the scientific community at large in a near future, and incorporate any feedback in its further improvements

ACKNOWLEDGMENTS

S. G. Djorgovski and C. Donalek acknowledge a partial support from the NSF grants HCC-0917814, IIS-1118041, and AST-1313422. Support for this work was provided in part by NASA through a contract issued by the Jet Propulsion Laboratory, California Institute of Technology under a contract with NASA. S. Davidoff is supported by NASA/JPL Raise-the-Bar and the Space Communication and Networking programs (SCAN). Djorgovski, Donalek, and Davidoff also acknowledge a partial support from a Caltech I-Grant. A. Cioc, J. Zhang, E. Lawler, and S. Yeh were supported in part by the Caltech SURF fellowships. A. Wang contributed to this work as a summer intern at Caltech.